\documentclass{article}
\usepackage{ijcai24}

\usepackage{times}
\usepackage{soul}
\usepackage{url}
\usepackage[hidelinks]{hyperref}
\usepackage[utf8]{inputenc}
\usepackage[small]{caption}
\usepackage{graphicx}
\usepackage{amsfonts}
\usepackage{amsmath}
\usepackage{amssymb}
\usepackage{amsthm}
\usepackage{siunitx}
\usepackage{booktabs}
\usepackage{algorithm}
\usepackage{algorithmic}
\usepackage{multirow}
\usepackage[switch]{lineno}
\usepackage[table]{xcolor}
\usepackage{colortbl}
\title{Separate in the Speech Chain: Cross-Modal Conditional Audio-Visual Target Speech Extraction}

\author{
Zhaoxi Mu
\and
Xinyu Yang\\
\affiliations
Xi'an Jiaotong University\\
\emails
wsmzxxh@stu.xjtu.edu.cn,
yxyphd@mail.xjtu.edu.cn
}

\begin{document}

\maketitle

\begin{abstract}
The integration of visual cues has revitalized the performance of the target speech extraction task, elevating it to the forefront of the field. Nevertheless, this multi-modal learning paradigm often encounters the challenge of modality imbalance. In audio-visual target speech extraction tasks, the audio modality tends to dominate, potentially overshadowing the importance of visual guidance. To tackle this issue, we propose AVSepChain, drawing inspiration from the speech chain concept. Our approach partitions the audio-visual target speech extraction task into two stages: speech perception and speech production. In the speech perception stage, audio serves as the dominant modality, while visual information acts as the conditional modality. Conversely, in the speech production stage, the roles are reversed. This transformation of modality status aims to alleviate the problem of modality imbalance. Additionally, we introduce a contrastive semantic matching loss to ensure that the semantic information conveyed by the generated speech aligns with the semantic information conveyed by lip movements during the speech production stage. Through extensive experiments conducted on multiple benchmark datasets for audio-visual target speech extraction, we showcase the superior performance achieved by our proposed method.
\end{abstract}

\section{Introduction}

Speech serves as the primary means of exchanging and communicating information among humans. However, in real-world settings, speech signals often encounter contamination from various interfering factors, such as noise and irrelevant speakers. This scenario is commonly known as the ``cocktail party problem" \cite{bronkhorst2000cocktail}. During actual conversations, perceivers possess the ability to selectively focus on specific acoustic stimuli, highlighting their inherent selective auditory attention. A long-term objective of speech processing research is to develop machines capable of emulating human auditory capabilities. These machines should be able to selectively extract the speech of a target speaker by utilizing auxiliary cues, such as reference speech \cite{XuRCL20,GeXWCD020,mu2023self}, static facial images \cite{GaoG21}, and lip movements \cite{WuXZCYXY19,GaoG21,PanTX021}.

Neuroscientific research has revealed that human perception of speech is inherently multi-modal, with lip movements being observed to enhance auditory perception during conversations \cite{li2018effects,stenzel2019limits}. Visual cues have proven to be robust auxiliary references for speaker extraction algorithms at the frame level, yielding remarkable outcomes in separating target speakers, particularly in the presence of highly overlapping speech \cite{EphratMLDWHFR18,OchiaiDKON19}. However, the utilization of static facial images has demonstrated limited improvement in the performance of target speech extraction (TSE) \cite{GaoG21} and may raise privacy concerns. Hence, this paper focuses on leveraging lip movements as the primary cue to guide TSE.

\begin{figure*}[tb]
    \centering
    \includegraphics[width=0.61\textwidth]{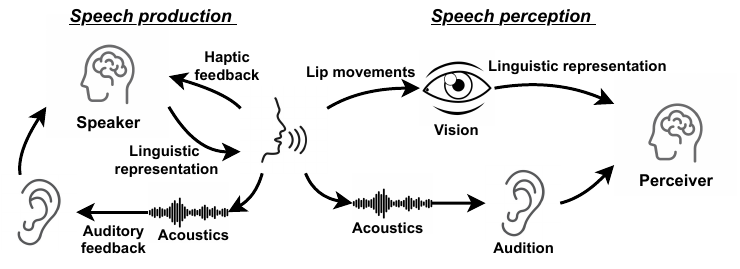}
    \caption{The schematic diagram of information flow within the speech chain. The speech chain captures the movement of information between the speaker and the perceiver during speech communication, encompassing the processes of speech production and speech perception.}
    \label{fig1}
\end{figure*}

In previous studies on audio-visual target speech extraction (AV-TSE), it has been observed that visual information is often underutilized, resulting in performance comparable to ignoring visual cues. This situation arises due to the audio-focused nature of AV-TSE, where the richness of visual feature information is less substantial compared to audio feature information. Consequently, the importance of visual cues has been insufficiently considered, creating an imbalance between audio and visual modalities. To address this issue, we draw inspiration from the concept of the speech chain \cite{denes1993speech} in speech science and propose the audio-visual speech separation chain (AVSepChain). The speech chain depicts the flow of information between speakers and perceivers in speech communication, encompassing two stages: speech production and speech perception \cite{deller1993discrete}. The information flow within the speech chain includes auditory feedback from the speaker's mouth to the ears, the perceiver receiving acoustic information from the speaker, and the perceiver receiving visual information regarding the speaker's lip movements, as shown in Figure \ref{fig1}.

In our approach, we employ a two-stage method to replicate the inverse process of the two stages of the speech chain. In the first stage, we utilize the audio-visual target speech extraction network to simulate speech perception. This network extracts the target speaker's speech from the mixed audio, leveraging the target speaker's lip movements as a guiding condition. In the second stage, we employ the lip-to-speech synthesis network to simulate speech production. This network generates speech based on lip movements, utilizing the speech extracted in the first stage as a condition. Unlike previous AV-TSE approaches that solely relied on the video modality for conditional guidance, our approach alleviates the modality imbalance issue by incorporating both audio and video modalities as mutual conditions.

To ensure the consistency between the speech content conveyed through the voice and lip movements of the target speaker, as depicted in the two channels of speech perception shown in Figure \ref{fig1}, we propose a contrastive semantic matching loss. Specifically, we employ audio and audio-visual self-supervised pre-trained models to extract frame-level quantized representations of speech content and lip movement content separately. These representations serve as pseudo-phonemes and pseudo-visemes, respectively, enabling us to align cross-modal semantics. A viseme represents a basic unit of visual speech, corresponding to a group of phonemes present in acoustic speech \cite{massaro2014speech,BearH17}. By capitalizing on the consistency between pseudo-visemes and pseudo-phonemes, we establish that the generated speech of the target speaker possesses the same semantic information as its lip movements.

In general, this paper presents three main contributions. (\romannumeral1) We introduce the concept of speech chain into AV-TSE, which, to the best of our knowledge, has not been explored previously. By leveraging the audio and video modalities as conditional information for each other, we simulate the speech perception and production processes in the speech chain, thereby alleviating the modality imbalance issue in AV-TSE. (\romannumeral2) We propose a contrastive semantic matching loss to regulate the generation of more precise audio signals in the lip-to-speech synthesis network. This loss conducts contrastive matching training on the content information of the audio and video modalities, which is extracted using pre-trained models. (\romannumeral3) We perform comprehensive experiments to validate the effectiveness of our approach. The results demonstrate that our proposed AVSepChain significantly enhances the state-of-the-art performance of AV-TSE.

\section{Related Work}

\subsection{Audio-Visual Target Speech Extraction}

Recent neuroscience research has demonstrated that the human brain effectively addresses the ``cocktail party problem" by incorporating visual cues \cite{li2018effects,stenzel2019limits}. Drawing on this understanding, researchers have integrated visual information into the paradigm of speech separation to enhance performance in noisy and challenging environments. Prior studies by \citeauthor{abs-2306-14170} \shortcite{abs-2306-14170} and \citeauthor{abs-2310-19581} \shortcite{abs-2310-19581} employed cross-modal attention for feature fusion, while \citeauthor{ZhouZ0Z00O22} \shortcite{ZhouZ0Z00O22} designed an adaptive multi-modal fusion framework. Additionally, \citeauthor{abs-2212-10744} \shortcite{abs-2212-10744} proposed a multi-scale fusion framework inspired by cortical-thalamic circuits. However, these methods typically prioritize audio as the dominant modality, resulting in an imbalanced treatment of audio and visual modalities.

\subsection{Lip-to-Speech Synthesis}
The tongue and lips assume a crucial role in controlling the shape of the vocal tract and modifying its resonance characteristics to convey distinct phonemes. Perceivers can infer a speaker's speech content by observing their lip movements, providing opportunities for integrating video into speech synthesis. Lip-to-speech synthesis (LTS) aims to convert visual information derived from the analysis of lip movements into corresponding speech signals. Similar to text-to-speech synthesis (TTS) approaches \cite{abs-2104-09995}, most LTS methods follow a two-stage procedure \cite{PrajwalMNJ20,KimHR21,MiraHPSP22,abs-2302-08841}, involving the generation of mel-spectrograms followed by their transformation into audio signals using vocoders. In this paper, we propose predicting the residual signal of the preliminary separated speech to alleviate the difficulty of the LTS task. Moreover, LTS inherently poses an underdetermined problem, as a single lip movement video can correspond to multiple speech possibilities. To impose constraints, previous research commonly introduced reference speech from the same speaker and employed a speaker encoder to obtain speaker embeddings as guiding information \cite{MiraHPSP22,abs-2305-19603,abs-2302-08841}. However, in this study, we utilize the target speaker's speech extracted during the speech perception stage as the registration, eliminating the need for additional speaker reference speech. Furthermore, we implicitly constrain the synthesis process by predicting the residual signal.
 
\subsection{Modality Imbalance Problem}

When dealing with audio-visual data, which involves multiple modalities, a common challenge is modality imbalance. Modality imbalance occurs when one modality dominates over others in a specific task or scenario. To alleviate this issue, a common approach is to adjust the training strategy of the model to balance the learning process for each modality. For instance, \citeauthor{Fan0WW023} \shortcite{Fan0WW023} introduced modality-specific loss functions and regularization terms to enhance the learning of non-dominant modalities. Similarly, \citeauthor{abs-2307-02041} \shortcite{abs-2307-02041} and \citeauthor{WuJCG22} \shortcite{WuJCG22} balanced the learning process of different modalities by dynamically adjusting gradients and learning rates, enabling the model to allocate more attention to non-dominant modalities. Another approach, as proposed by \citeauthor{abs-2106-11059} \shortcite{abs-2106-11059}, involves leveraging pre-trained single-modal models to tackle the modality imbalance problem. They extracted features from the single-modal model and distilled them into the multi-modal model, thereby improving the learning of non-dominant modalities. In this study, we alleviate the issue of modality imbalance by shifting the dominant role between two modalities at different stages.

\section{Methodology}

\begin{figure*}[tb]
    \centering
    \includegraphics[width=0.92\textwidth]{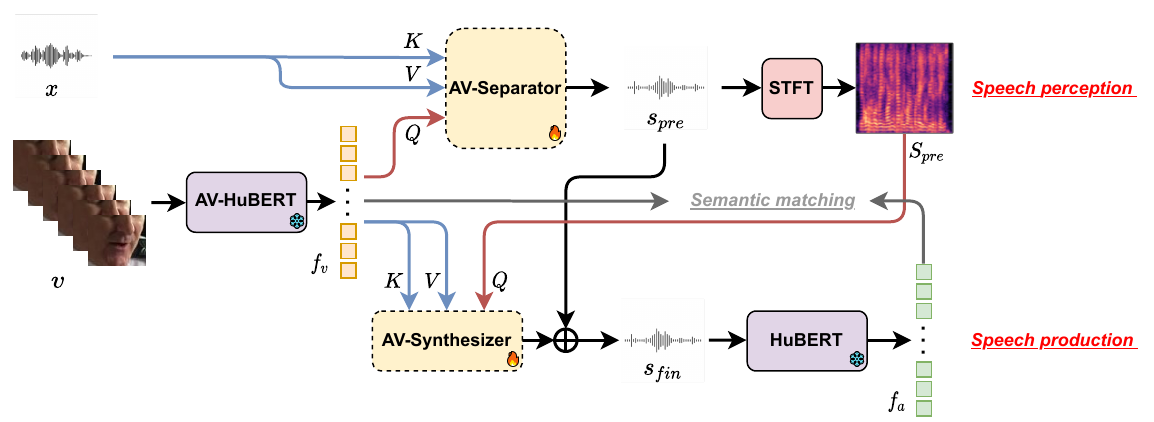}
    \caption{The overall framework of AVSepChain encompasses two stages: speech perception and speech production. In the speech perception stage, the AV-Separator initially extracts the target speaker's speech. In the speech production stage, the AV-Synthesizer predicts the residual signal of the output from the speech perception stage. In the speech perception stage, audio is treated as the dominant modality, while visual information serves as the conditional modality. This relationship is reversed in the speech production stage. AV-HuBERT and HuBERT, depicted in the solid line box, have their parameters fixed during training. The AV-Separator and AV-Synthesizer, shown in the dotted box, have their parameters updated during training. The embeddings extracted by AV-HuBERT and HuBERT are utilized to calculate the contrastive modality matching loss.}
    \label{fig2}
\end{figure*}

\subsection{Overview} 

When presented with a mixed speech signal $x$ and a lip movement video $v$ of the target speaker, the proposed AVSepChain framework aims to extract clean speech $s$ corresponding to the target speaker from the speech of interfering speakers and environmental noise by leveraging the visual information. The framework consists of two stages: speech perception and speech production, as illustrated in Figure \ref{fig2}. In the speech perception stage, we employ AV-HuBERT \cite{ShiHLM22}, a pre-trained audio-visual model, to extract the visual hidden unit $f_v$ from the lip movement video $v$. Subsequently, the audio-visual speech separator (AV-Separator) is utilized to preliminarily separate the target speech $s_{pre}$ from the mixed speech $x$, using $f_v$ as a condition. 

In the speech production stage, the mel-spectrogram $S_{pre}$ of the preliminary separated target speech $s_{pre}$ obtained in the first stage serves as a conditional input. The audio-visual speech synthesizer (AV-Synthesizer) predicts the residual signal $s_{res}$ of the target speech based on the visual hidden unit $f_v$. The final target speech $s_{fin}$ is obtained by adding $s_{pre}$ and $s_{res}$. Moreover, we utilize HuBERT \cite{HsuBTLSM21}, a pre-trained audio model, to extract the audio hidden unit $f_a$ from $s_{fin}$. This unit is matched with $f_v$ to ensure semantic feature alignment, thus guaranteeing that the generated target speech captures the same semantic information as the lip movement. Each section will be further elaborated in the subsequent sections.

\subsection{Speech Perception}

\paragraph{AV-HuBERT.} In the speech perception stage, AV-HuBERT functions as the encoder for lip movement videos. AV-HuBERT is a self-supervised model designed for audio-visual learning, which predicts clustering assignments in masked regions based on speech features and lip movement sequences. Previous studies have demonstrated the efficacy of this approach in capturing meaningful semantic information conveyed by lip movements. Consequently, it offers significant benefits for downstream tasks such as audio-visual speech recognition \cite{ShiHLM22,interspeech/ShiHM22}.

In our approach, for the lip movement video of the target speaker, represented as $v \in \mathbb{R}^{H\times W\times T_v}$, where $H$, $W$, and $T_v$ denote the height, width, and number of image frames respectively, we input it into the pre-trained AV-HuBERT. This results in a condensed frame-level visual representation, denoted as $f_v \in \mathbb{R}^{N_{f_v}\times T_{f_v}}$, where $N_{f_v}$ signifies the number of channels of $f_v$, and $T_{f_v}$ indicates its length.

\paragraph{AV-Separator.} In our approach, the AV-Separator utilizes the widely adopted AV-Sepformer \cite{abs-2306-14170} as its backbone. Sepformer \cite{SubakanRCBZ21} is a time-domain monaural speech separation model that leverages the dual-path Transformer architecture. AV-Sepformer extends Sepformer by incorporating the cross-attention mechanism \cite{VaswaniSPUJGKP17} to integrate both audio and visual modalities. In this study, similar to AV-Sepformer, we employ cross-modal attention to fuse audio and video features. In this case, considering that the visual information can be regarded as cues for separating the target speech, we adopt an approach where the audio modality is treated as the dominant modality while the visual modality acts as the conditional modality.

Specifically, we initially encode the mixed audio $x \in \mathbb{R}^{T_a}$ through a 1-D convolutional encoder, resulting in an audio feature representation $X \in \mathbb{R}^{N_a\times T_X}$. Here, $N_a$ represents the number of convolutional kernels, and $T_X$ denotes the length of the audio feature. Subsequently, we divide the 2-D audio feature $X$ into chunks of length $K_X$, with a stride size of $\frac{K_X}{2}$. These chunks are then concatenated to form a 3-D audio feature $X_c \in \mathbb{R}^{N_a\times K_X\times S_X}$, where $S_X$ refers to the number of chunks. To match the dimensions of the video feature $f_v$ and the audio feature $X_c$, we replicate $f_v$ $K_X$ times, adjust the length of $f_v$ to $S_X$ using interpolation, and project the channel dimension of $f_v$ to $N_a$, resulting in the adjusted video feature $\tilde{f}_v \in \mathbb{R}^{N_a\times K_X\times S_X}$.

Next, we utilize the visual feature $\tilde{f}_v$ as the query and the audio feature $X_c$ as the key and value for cross-modal attention fusion. This fusion yields an audio-dominant multi-modal feature denoted as $H_{av} \in \mathbb{R}^{N_a\times K_X\times S_X}$, which can be computed as:
\begin{equation}
H_{av}=\text{Softmax} \left( \frac{\tilde{f}_v W_v^Q \cdot \left(X_c W_a^K\right)^\intercal}{\sqrt{N_a}} \right) X_c W_a^V
\end{equation}
Here, $W_v^Q$, $W_a^K$, and $W_a^V \in \mathbb{R}^{N_a \times N_a}$ are learnable parameters. Subsequently, the AV-Separator extracts the speech of the target speaker $s_{pre} \in \mathbb{R}^{T_a}$ from the audio-dominant cross-modal feature $H_{av}$. To accomplish this, we utilize the scale-invariant signal-to-noise ratio (SI-SNR) \cite{RouxWEH19} between $s_{pre}$ and the true signal as the loss function $\mathcal{L}_{\text{per}}$ during the speech perception stage. This loss function aims to separate the speech of the target speaker preliminarily.

\subsection{Speech Production}

\paragraph{AV-Synthesizer.} During the speech production stage, it is crucial to handle the intrinsic one-to-many relationship between visemes and phonemes, as well as accommodate speech-related variations such as voice, accents, and prosody. To tackle these challenges, we leverage the target speech that was extracted in the previous stage as a registration to facilitate the synthesis of speech from lip movements. In contrast to the previous stage, we assign vision as the dominant modality and audio as the conditional modality in this stage, aiming to mitigate modality imbalance issues.

Specifically, we utilize the mel-spectrogram $S_{pre} \in \mathbb{R}^{N_{mel} \times T_{mel}}$ of the output $s_{pre}$ from the first stage, along with the lip movement feature sequence $f_v$ extracted by AV-HuBERT, as the audio-visual input to the AV-Synthesizer. Here, $N_{mel}$ denotes the channel number of $S_{pre}$, and $T_{mel}$ represents its length. Since the video stream and the audio stream are inherently synchronized, the visual features and the mel-spectrogram are naturally aligned. Consequently, we upsample the video frames to match the sampling rate of the mel-spectrogram based on the ratio between the sampling rates of the mel-spectrogram and the video frames. Additionally, we project the video feature $f_v$ and audio feature $S_{pre}$ to the same channel dimension $N_{pro}$, resulting in $\hat{f}_v$ and $\hat{S}_{pre} \in \mathbb{R}^{N_{pro} \times T_{mel}}$. 

Next, we employ the audio feature $\hat{S}_{pre}$ as the query and the visual feature $\hat{f_v}$ as the key and value for cross-modal attention fusion. This fusion yields a visual-dominant multi-modal feature, denoted as $H_{va} \in \mathbb{R}^{N_{pro} \times T_{mel}}$, which can be expressed as: 
\begin{equation}
H_{va}=\text{Softmax} \left( \frac{ \hat{S}_{pre} W_a^Q \cdot \left( \hat {f}_v W_v^K \right) ^\intercal}{\sqrt{N_{pro}}} \right) \hat{f}_v W_v^V
\end{equation}
Here, $W_a^Q$, $W_v^K$, and $W_v^V \in \mathbb{R}^{N_{pro}\times N_{pro}}$ represent the learnable parameters. Subsequently, the AV-Synthesizer utilizes the visual-dominant multi-modal feature $H_{va}$ to predict the residual signal $s_{res} \in \mathbb{R}^{T_a}$ of the target speech $s_{pre}$, which was extracted in the first stage. The final generated speech of the target speaker $s_{fin} \in \mathbb{R}^{T_a}$ is obtained by adding the residual signal $s_{res}$ to the preliminary result $s_{pre}$, as shown in the equation below:
\begin{equation}
s_{fin}=s_{pre}+s_{res}
\end{equation}

During the speech production stage, our approach involves predicting the residual signal rather than the complete speech waveform. This strategy aims to provide a preliminary candidate for the AV-Synthesizer, thus alleviating the challenge of modeling the entire speech signal. By focusing on the residuals, the AV-Synthesizer can effectively capture the remaining information necessary for synthesis and refine the preliminary results. This strategy also incorporates an implicit constraint into the synthesis process. To ensure that the generated speech accurately corresponds to lip movements, we also employ the SI-SNR between $s_{fin}$ and the true signal as the loss function $\mathcal{L}_{\text{syn}}$ during the speech production stage.

\begin{table*}[th]
\centering
\small
\setlength\tabcolsep{3pt}  
\renewcommand{\arraystretch}{1.2} % 控制行间距的倍数
\begin{tabular}{ccccccccc}
\hline
\multirow{2}{*}{Model} & \multicolumn{4}{c}{LRS2-2Mix} & \multicolumn{4}{c}{VoxCeleb2-2mix} \\ \cline{2-9} 
                       & SI-SNRi$\uparrow$  & SDRi$\uparrow$ & PESQ$\uparrow$ & WER (\%)$\downarrow$  & SI-SNRi$\uparrow$   & SDRi$\uparrow$   & PESQ$\uparrow$  & WER (\%)$\downarrow$   \\ \hline
AV-ConvTasNet \shortcite{WuXZCYXY19}          & 12.4     & 12.7 & 2.75 & 31.4 & 10.6      & 10.9   & 2.07  & 34.2  \\
Visualvoice \shortcite{GaoG21}           & 11.5     & 11.8 & 3.00 & 34.5 & 9.3       & 10.2   & 1.97  & 36.4  \\
MuSE \shortcite{PanTX021}                  & 13.5     & 13.8 & 2.97 & 27.9 & 11.7      & 12.0   & 2.21  & 32.0  \\
CTCNet \shortcite{abs-2212-10744}                & 14.3     & 14.6 & 3.06 & 24.8 & 11.9      & 13.1   & 2.26  & 26.8  \\
AV-SepFormer \shortcite{abs-2306-14170}          & 14.1     & 14.4 & 3.15 & 25.4 & 12.1      & 12.5   & 2.31  & 27.3  \\
\rowcolor{gray!20}AVSepChain (Ours)      & \textbf{15.3}     & \textbf{15.7} & \textbf{3.26} & \textbf{20.2} & \textbf{13.6}      & \textbf{14.2}   & \textbf{2.72}  & \textbf{22.1}  \\ \hline
\end{tabular}
\caption{Performance comparison of our method with state-of-the-art AV-TSE methods on the LRS2-2Mix and VoxCeleb2-2mix datasets. We have referenced the original literature for results that are already established. For outcomes not documented in the original works, we have incorporated findings from our own replication efforts.}
\label{tab1}
\end{table*}

\begin{table}[th]
\centering
\small
\setlength\tabcolsep{2pt}  
\renewcommand{\arraystretch}{1.2} % 控制行间距的倍数
\begin{tabular}{ccccc}
\hline
\multirow{2}{*}{Model} & \multicolumn{2}{c}{LRS3} & \multicolumn{2}{c}{TCD-TIMIT} \\ \cline{2-5} 
                       & SI-SNRi$\uparrow$      & PESQ$\uparrow$      & SI-SNRi$\uparrow$         & PESQ$\uparrow$        \\ \hline
AV-ConvTasNet \shortcite{WuXZCYXY19}         & 12.1         & 2.33      & 11.5            & 2.21        \\
Visualvoice \shortcite{GaoG21}           & 11.6         & 2.27      & 10.9            & 2.25        \\
MuSE \shortcite{PanTX021}                  & 13.0         & 2.56      & 12.5            & 2.45        \\
AV-SepFormer \shortcite{abs-2306-14170}           & 13.8         & 2.67      & 13.4            & 2.57        \\
\rowcolor{gray!20}AVSepChain (Ours)      & \textbf{15.2} & \textbf{3.12} & \textbf{14.7} & \textbf{2.88}        \\ \hline
\end{tabular}
\caption{Cross-domain performance comparison by training on the VoxCeleb2-2Mix dataset and testing on the LRS3 and TCD-TIMIT datasets.}
\label{tab1.5}
\end{table}

\begin{table}[th]
\centering
\small
\setlength\tabcolsep{2pt}  
\renewcommand{\arraystretch}{1.2}
\begin{tabular}{ccccc}
\hline
Method             &  SI-SNRi$\uparrow$ & SDRi$\uparrow$ & PESQ$\uparrow$ & WER (\%)$\downarrow$  \\ \hline
\rowcolor{gray!20}AVSepChain         & \textbf{15.3} & \textbf{15.7}    & \textbf{3.26} & \textbf{20.2} \\
w/o AV-Synthesizer & 14.2 & 14.5   & 3.13 & 26.0 \\ 
w/o $\mathcal{L}_{\text{mat}}$  & 14.5 & 14.8    & 3.17 & 24.1 \\ 
Predict complete signal  & 14.3 & 14.6    & 3.16 & 25.8 \\
\hline
\end{tabular}
\caption{Ablation study on the speech production stage, contrastive semantic matching loss, and prediction of residual signal on the LRS2-2Mix dataset.}
\label{tab2}
\end{table}

\begin{table*}[th]
\centering
\small
\renewcommand{\arraystretch}{1.2}
\setlength\tabcolsep{3pt}   
\begin{tabular}{cccccc}
\hline
Network Input & Calculating Loss & SI-SNRi$\uparrow$          & SDRi$\uparrow$       & PESQ$\uparrow$          & WER (\%)$\downarrow$           \\ \hline
\rowcolor{gray!20}AV-HuBERT     & AV-HuBERT        & \textbf{15.3} & \textbf{15.7} & \textbf{3.26} & \textbf{20.2} \\
ResNet-18     & AV-HuBERT        & 14.5          & 14.8          & 3.08          & 23.5          \\
MS-TCN        & AV-HuBERT        & 14.6          & 14.9          & 3.12          & 22.1          \\
CTCNet-Lip    & AV-HuBERT        & 14.9          & 15.3          & 3.24          & 21.8          \\
AV-HuBERT     & ResNet-18        & 14.6          & 14.9          & 3.10          & 24.9          \\
AV-HuBERT     & MS-TCN           & 14.9          & 15.3          & 3.14          & 24.2          \\
AV-HuBERT     & CTCNet-Lip       & 15.1          & 15.4          & 3.21          & 23.8          \\ \hline
\end{tabular}
\caption{Results obtained employing visual embeddings extracted by various visual front-ends on the LRS2-2Mix dataset. These embeddings serve as input for the AV-Separator and AV-Synthesizer, as well as for calculating the contrastive semantic matching loss.}
\label{tab3}
\end{table*}

\begin{table}[th]
\centering
\small
\renewcommand{\arraystretch}{1.2}
\setlength\tabcolsep{3pt}   
\begin{tabular}{ccccc}
\hline
Method          & SI-SNRi$\uparrow$       & SDRi$\uparrow$          & PESQ$\uparrow$          & WER (\%)$\downarrow$           \\ \hline
\rowcolor{gray!20}Cross-attention & \textbf{15.3} & \textbf{15.7} & \textbf{3.26} & \textbf{20.2} \\
Concatenation   & 13.5          & 13.8          & 3.06          & 26.7          \\
Summation       & 13.2          & 13.6          & 2.98          & 28.3          \\ \hline
\end{tabular}
\caption{Results of employing various cross-modal modulation strategies on the LRS2-2Mix dataset for both the AV-Separator and AV-Synthesizer.}
\label{tab4}
\end{table}

\begin{table*}[th]
\centering
\small
\renewcommand{\arraystretch}{1.2}
\setlength\tabcolsep{4pt}   
\begin{tabular}{cccccccc}
\hline
\multicolumn{2}{c}{Speech Perception} & \multicolumn{2}{c}{Speech Production} & \multirow{2}{*}{SI-SNRi$\uparrow$} & \multirow{2}{*}{SDRi$\uparrow$} & \multirow{2}{*}{PESQ$\uparrow$} & \multirow{2}{*}{WER (\%)$\downarrow$} \\ \cline{1-4}
Query          & Key \& Value         & Query          & Key \& Value         &                       &                          &                       &                      \\ \hline
\rowcolor{gray!20}Video          & Audio                & Audio          & Video                & \textbf{15.3}         & \textbf{15.7}            & \textbf{3.26}         & \textbf{20.2}        \\
Audio          & Video                & Audio          & Video                & 14.6                  & 14.9                     & 3.03                  & 24.0                 \\
Video          & Audio                & Video          & Audio                & 13.8                  & 14.1                     & 2.86                  & 28.4                 \\
Audio          & Video                & Video          & Audio                & 13.6                  & 13.9                     & 2.82                  & 29.3                 \\ \hline
\end{tabular}
\caption{Results of utilizing different modalities as the query, key, and value in the cross-modal attention mechanism during the speech perception and speech production stages on the LRS2-2Mix dataset.}
\label{tab5}
\end{table*}

\paragraph{Semantic matching.} To ensure that the speech generated by the AV-Synthesizer conveys the same semantic information as the corresponding lip movements, we propose a contrastive semantic matching loss. Initially, we employ a pre-trained HuBERT to extract the frame-level speech representation $f_a \in \mathbb{R}^{N_{f_a} \times T_{f_a}}$ from the generated speech $s_{fin}$. Here, $N_{f_a}$ and $T_{f_a}$ denote the channel number and length of $f_a$, respectively. It is worth mentioning that this representation has demonstrated effectiveness in downstream tasks like automatic speech recognition (ASR) \cite{HsuBTLSM21,ShiHM22}. Therefore, $f_a$ can be considered as a pseudo-phoneme representation. Similarly, the representation $f_v$ extracted by AV-HuBERT can be regarded as a pseudo-viseme representation. To align the cross-modal semantic representations, we introduce a contrastive semantic matching loss, denoted as $\mathcal{L}_{\text{mat}}$, which is defined as follows:
\begin{equation}
\mathcal{L}_{\text{mat}}=\text{max} \{d(n(f_v),n(f_a))- d(n(f_v),n(\bar{f}_a))+m,0\}
\end{equation}
Here, $d( \cdot, \cdot)$ represents the L2 distance, $n$ denotes L2 normalization, $\bar{f_a}$ is the embedding obtained by applying HuBERT to the interfering audio $x-s_{fin}$, and $m$ represents the margin. To compute $\mathcal{L}_{\text{mat}}$, we index each frame of the pseudo-phoneme representations to their corresponding pseudo-viseme representations based on the ratio of the sampling rates of the audio and video streams. This ensures that the lengths of $f_a$, $\bar{f}_a$, and $f_v$ are matched. The gradient of $\mathcal{L}_{\text{mat}}$ is then back-propagated to the AV-Synthesizer to update its parameters.

\subsection{Training}
The overall training objective of our method can be defined as follows:
\begin{equation}
\mathcal{L}_{\text{total}} = \mathcal{L}_{\text{per}} + \mathcal{L}_{\text{syn}} + \lambda \mathcal{L}_{\text{mat}} 
\end{equation}
Here, $\lambda$ is a weighting factor. The loss function SI-SNR, which is used to calculate $\mathcal{L}_{\text{per}}$ and $\mathcal{L}_{\text{syn}}$, is defined as:
\begin{equation}
\text{SI-SNR}(u,\hat{u})=-10\log_{10} \left( \frac{\lVert \frac{\langle \hat{u},u \rangle u}{\lVert u \rVert^2} \rVert^2}{\lVert \hat{u}- \frac{\langle \hat{u},u \rangle u}{\lVert u \rVert^2} \rVert^2} \right)
\end{equation}
In the above equations, $u$ and $\hat{u}$ represent the ground truth and predicted speech of the target speaker, respectively. $\lVert u \rVert^2 = \langle u,u \rangle$ denotes the signal power. During training, we optimize the parameters of the AV-Separator and AV-Synthesizer while keeping the parameters of AV-HuBERT and HuBERT fixed.

\section{Experiments}

\subsection{Datasets}
We conducted experiments on two audio-visual datasets, namely LRS2-2Mix and VoxCeleb2-2Mix \cite{abs-2212-10744}, derived from the LRS2 \cite{AfourasCSVZ22} and VoxCeleb2 \cite{ChungNZ18} datasets, respectively. Both LRS2-2Mix and VoxCeleb2-2Mix consist of mixtures of two speakers. We randomly selected two distinct speakers and mixed their speech with a signal-to-noise ratio ranging from -5 dB to 5 dB. These datasets contain multilingual speech corrupted by noise. The speaker identities in the training and test sets do not overlap. Each speech segment has a duration of 2 seconds and a sampling rate of 16 kHz. The video frames are synchronized with the speech at a frame rate of 25 FPS. In our experiments, we cropped the lip region and resized the cropped frames to $88 \times 88$ pixels, resulting in a height and width of $H=W=88$. 

\subsection{Implementation Details}

For both AV-HuBERT and HuBERT, we employ a 12-layer BASE pre-trained model with a feature dimension of 768, denoted as $N_{f_v}=N_{f_a}=768$, and extract the features from the last layer as embeddings. The AV-Separator is built on the same hyperparameters as the AV-Sepformer \cite{abs-2306-14170}, comprising two repetitions of 8 Intra-Transformers, 7 Inter-Transformers, and 1 Cross-Modal Transformer. The values for $N_a$ and $K_X$ are set to 256 and 160, respectively. We employ a logarithmic mel-spectrogram with 80 mel bands, a filter length of 1024, a hop size of 10 ms, a window length of 40 ms, and a Hann window to capture the spectro-temporal features $S_{pre}$ of the output audio $s_{pre}$ from the AV-Separator. In other words, $N_{mel}$ is 80. The intermediate feature dimension $N_{pro}$ is set to 256. The AV-Synthesizer consists of a cross-modal attention layer, followed by three 1-D convolution layers. The hidden dimensions for the convolution layers are 256, 128, and 160, respectively, with a kernel size of 7. Overall, our model encompasses a total of 33.1M trainable parameters, with 31.3M parameters in the AV-Separator and 1.8M parameters in the AV-Synthesizer. When calculating the contrastive semantic matching loss, we set the margin $m$ to 0.5 and $\lambda$ to 1. The model is trained using Adam optimization, starting with an initial learning rate of $1.5 \times 10^{-4}$. If the loss does not decrease on the validation set for three consecutive epochs, the learning rate is halved. The training process is terminated if there is no decrease in the loss for five consecutive epochs.

\subsection{Evaluation Metrics}

We evaluate the quality of separated speech using scale-invariant signal-to-noise ratio improvement (SI-SNRi) \cite{RouxWEH19} and signal-to-noise ratio improvement (SDRi) \cite{VincentGF06}. The overall perceptual quality is measured using the perceptual evaluation of speech quality (PESQ) \cite{RixBHH01}. Moreover, to assess whether the proposed method, as a speech enhancement front-end, improves the accuracy of downstream tasks such as ASR, we utilize the publicly available Google speech-to-text API to obtain recognition results. Our primary focus lies on measuring the word error rate (WER), where a lower error rate indicates better preservation and restoration of the content information in the speech.

\subsection{Comparison with the State-of-the-Art}

We comprehensively compared AVSepChain and existing AV-TSE methods on the LRS2-2Mix and VoxCeleb2-2Mix datasets. The results, presented in Table \ref{tab1}, demonstrate that AVSepChain achieves state-of-the-art performance on both datasets. Notably, our proposed AVSepChain significantly outperforms other methods in terms of PESQ and WER metrics. These findings suggest that our method, designed to simulate the perception and production processes of the speech chain, enhances the perceptual quality of separated speech and can be effectively integrated into downstream ASR tasks.

\paragraph{Cross-domain performance comparison.} To assess the generalization performance of our approach, we conducted cross-domain performance tests on additional audio-visual datasets. Specifically, we trained our model using the VoxCeleb2-2Mix dataset and evaluated its performance on the LRS3 \cite{abs-1809-00496} and TCD-TIMIT \cite{HarteG15} datasets, derived from TED videos and studio recordings, respectively. For these datasets, we randomly selected speech segments from two distinct speakers and mixed them, following a similar approach as in \cite{PanTX021,abs-2306-14170}. The results, presented in Table \ref{tab1.5}, demonstrate that our proposed method exhibits robust generalization capabilities despite potential distribution shifts, which are of utmost importance in real-world applications.

\subsection{Ablation Study}

In this section, we performed ablation experiments to validate the effectiveness of each key design proposed in AVSepChain. All experiments were conducted on the LRS2-2Mix dataset.

\paragraph{Ablation study on speech production.} To investigate the impact of the speech production process on the separation performance in AVSepChain, we excluded the AV-Synthesizer and utilized the output of the AV-Separator as the final separated speech. The results are presented in the ``w/o AV-Synthesizer" row of Table \ref{tab2}. These findings reveal that the speech production stage significantly enhances the performance of the AV-TSE task by compensating for the limitations of speech perception.

\paragraph{Ablation study on semantic matching.} To examine the impact of the contrastive semantic matching loss, we excluded this component during training and solely utilized the SI-SNR loss to train AVSepChain. The findings are presented in the ``w/o $\mathcal{L}_{\text{mat}}$" row of Table \ref{tab2}. The notable enhancements observed in the PESQ and WER metrics emphasize the vital role played by the contrastive semantic matching loss in facilitating the AV-Synthesizer to learn accurate pronunciation and speech content.

\paragraph{Ablation study on prediction of residual signal.} During the speech production stage, we compensate for the target speech, extracted during the speech perception stage, by predicting the residual signal. To emphasize the advantage of this strategy, we compare the results achieved by predicting the complete speech signal, as indicated in the ``Predict complete signal" row of Table \ref{tab2}. The findings demonstrate a significant improvement in speech production through the prediction of the residual signal. This improvement can be attributed to the ill-posed nature of the lip-to-speech synthesis task, which encounters a one-to-many problem. By focusing on predicting the residual signal, we effectively reduce the complexity of the task and introduce implicit constraints into the synthesis process, leading to enhanced generation results of the AV-Synthesizer with a smaller scale. In future research, we aim to explore the utilization of powerful generative models, such as diffusion models \cite{abs-2306-17203,abs-2308-07787}, to directly generate complete speech signals during the speech production stage and improve the upper limit of our method's performance \cite{abs-2301-10752}.

\paragraph{Comparison of visual front-ends.} To validate the superiority of AV-HuBERT as the visual front-end, we compared it with several popular pre-trained lip-reading models, including ResNet-18 \cite{AfourasCZ18}, MS-TCN \cite{Martinez0PP20}, and CTCNet-Lip \cite{abs-2212-10744}. In this experiment, we employed these models to extract visual embeddings, which were then utilized as inputs for the AV-Separator and AV-Synthesizer, as well as for calculating the contrastive semantic matching loss. The experimental results, presented in Table \ref{tab3}, consistently reveal the superior performance of AV-HuBERT over other models, both as input to the network and for computing the loss. Notably, AV-HuBERT significantly outperforms other pre-trained lip-reading models as a pseudo-viseme extractor in estimating the semantic matching loss. This superior performance can be attributed to the integration of visual (lip movement) and auditory (audio signals) information in AV-HuBERT, which enables the model to grasp the relationship between lip movement and speech, thereby enhancing lip-reading accuracy. Moreover, AV-HuBERT shares a similar self-supervised learning framework with HuBERT, involving feature clustering and masked prediction. Consequently, utilizing AV-HuBERT as a pseudo-viseme extractor leads to semantic feature representations that better align with those of HuBERT, resulting in improved effectiveness of the contrastive semantic matching loss.

\paragraph{Comparison of modulation strategies.} To validate the effectiveness of cross-modal modulation using cross-modal attention, we compared various modulation strategies, including concatenation and summation. The results are presented in Table \ref{tab4}. The findings demonstrate a significant superiority of the employed cross-modal attention method compared to other approaches. This superiority stems from the adaptive learning capability of cross-modal attention, which enables the determination of weights for each modality rather than simply feature combination or addition. Consequently, the model can accurately control the contribution of each modality by dynamically adjusting the dominant and conditional status of different modalities according to the specific task requirements, achieved by modifying the modalities of the query, key, and value. Furthermore, the model can selectively focus on the most pertinent and valuable information within each modality based on the specific task, thereby reducing the impact of redundant information. This refinement enhances the robustness and generalization ability of our model.

\paragraph{The impact of changing dominant and conditional modalities.} In AVSepChain, we alleviate the issue of modality imbalance by dynamically switching between dominant and conditional modalities. This is accomplished by modifying the queries, keys, and values of the cross-modal attention in various tasks. To assess the effectiveness of our approach, we conducted comparative experiments by altering the order of queries, keys, and values. The results of these experiments are presented in Table \ref{tab5}. As the keys and values in the cross-modal attention should originate from the same feature, there are four potential settings. The outcomes demonstrate that our configuration consistently achieves the most favorable performance. Additionally, the table reveals that using either audio or visual modalities as the dominant modality for both the speech perception and speech production stages results in a significant modality imbalance problem, potentially leading to failure in the other conditional modality. When the positions of the audio and visual modalities are swapped, the negative impact increases considerably, suggesting that the dominant modality should be assigned as the keys and values, while the conditional modality should serve as the query.

\section{Conclusion}

From the perspective of the speech chain, this paper proposes AVSepChain, which simulates the process of speech perception and speech production in the speech chain. Its objective is to alleviate the modality imbalance problem in AV-TSE by considering audio and visual modalities as conditional modalities for each other. This is achieved by assigning different modalities as queries, keys, and values for cross-modal attention. Furthermore, to ensure that the synthesized speech preserves the same semantic information as the lip movement video during the speech production process, a contrastive semantic matching loss is proposed. Extensive experiments confirm the effectiveness of our approach, indicating that our method defines state-of-the-art performance on the AV-TSE task by mitigating modality imbalance.

\bibliographystyle{named}
\bibliography{ijcai24}

\begin{thebibliography}{}

\bibitem[\protect\citeauthoryear{Afouras \bgroup \em et al.\egroup }{2018a}]{AfourasCZ18}
Triantafyllos Afouras, Joon~Son Chung, and Andrew Zisserman.
\newblock The conversation: Deep audio-visual speech enhancement.
\newblock In {\em {INTERSPEECH}}, pages 3244--3248. {ISCA}, 2018.

\bibitem[\protect\citeauthoryear{Afouras \bgroup \em et al.\egroup }{2018b}]{abs-1809-00496}
Triantafyllos Afouras, Joon~Son Chung, and Andrew Zisserman.
\newblock {LRS3-TED:} a large-scale dataset for visual speech recognition.
\newblock {\em CoRR}, abs/1809.00496, 2018.

\bibitem[\protect\citeauthoryear{Afouras \bgroup \em et al.\egroup }{2022}]{AfourasCSVZ22}
Triantafyllos Afouras, Joon~Son Chung, Andrew~W. Senior, Oriol Vinyals, and Andrew Zisserman.
\newblock Deep audio-visual speech recognition.
\newblock {\em {IEEE} Trans. Pattern Anal. Mach. Intell.}, 44(12):8717--8727, 2022.

\bibitem[\protect\citeauthoryear{Bear and Harvey}{2017}]{BearH17}
Helen~L. Bear and Richard~W. Harvey.
\newblock Phoneme-to-viseme mappings: the good, the bad, and the ugly.
\newblock {\em Speech Commun.}, 95:40--67, 2017.

\bibitem[\protect\citeauthoryear{Bronkhorst}{2000}]{bronkhorst2000cocktail}
Adelbert~W Bronkhorst.
\newblock The cocktail party phenomenon: A review of research on speech intelligibility in multiple-talker conditions.
\newblock {\em Acta Acustica united with Acustica}, 86(1):117--128, 2000.

\bibitem[\protect\citeauthoryear{Choi \bgroup \em et al.\egroup }{2023a}]{abs-2308-07787}
Jeongsoo Choi, Joanna Hong, and Yong~Man Ro.
\newblock Diffv2s: Diffusion-based video-to-speech synthesis with vision-guided speaker embedding.
\newblock {\em CoRR}, abs/2308.07787, 2023.

\bibitem[\protect\citeauthoryear{Choi \bgroup \em et al.\egroup }{2023b}]{abs-2305-19603}
Jeongsoo Choi, Minsu Kim, and Yong~Man Ro.
\newblock Intelligible lip-to-speech synthesis with speech units.
\newblock {\em CoRR}, abs/2305.19603, 2023.

\bibitem[\protect\citeauthoryear{Chung \bgroup \em et al.\egroup }{2018}]{ChungNZ18}
Joon~Son Chung, Arsha Nagrani, and Andrew Zisserman.
\newblock Voxceleb2: Deep speaker recognition.
\newblock In {\em {INTERSPEECH}}, pages 1086--1090. {ISCA}, 2018.

\bibitem[\protect\citeauthoryear{de Mira \bgroup \em et al.\egroup }{2022}]{MiraHPSP22}
Rodrigo Schoburg~Carrillo de~Mira, Alexandros Haliassos, Stavros Petridis, Bj{\"{o}}rn~W. Schuller, and Maja Pantic.
\newblock {SVTS:} scalable video-to-speech synthesis.
\newblock In {\em {INTERSPEECH}}, pages 1836--1840. {ISCA}, 2022.

\bibitem[\protect\citeauthoryear{Deller~Jr}{1993}]{deller1993discrete}
John~R Deller~Jr.
\newblock Discrete-time processing of speech signals.
\newblock In {\em Discrete-time processing of speech signals}, pages 908--908. 1993.

\bibitem[\protect\citeauthoryear{Denes and Pinson}{1993}]{denes1993speech}
Peter~B Denes and Elliot Pinson.
\newblock {\em The speech chain}.
\newblock Macmillan, 1993.

\bibitem[\protect\citeauthoryear{Du \bgroup \em et al.\egroup }{2021}]{abs-2106-11059}
Chenzhuang Du, Tingle Li, Yichen Liu, Zixin Wen, Tianyu Hua, Yue Wang, and Hang Zhao.
\newblock Improving multi-modal learning with uni-modal teachers.
\newblock {\em CoRR}, abs/2106.11059, 2021.

\bibitem[\protect\citeauthoryear{Ephrat \bgroup \em et al.\egroup }{2018}]{EphratMLDWHFR18}
Ariel Ephrat, Inbar Mosseri, Oran Lang, Tali Dekel, Kevin Wilson, Avinatan Hassidim, William~T. Freeman, and Michael Rubinstein.
\newblock Looking to listen at the cocktail party: a speaker-independent audio-visual model for speech separation.
\newblock {\em {ACM} Trans. Graph.}, 37(4):112, 2018.

\bibitem[\protect\citeauthoryear{Fan \bgroup \em et al.\egroup }{2023}]{Fan0WW023}
Yunfeng Fan, Wenchao Xu, Haozhao Wang, Junxiao Wang, and Song Guo.
\newblock {PMR:} prototypical modal rebalance for multimodal learning.
\newblock In {\em {CVPR}}, pages 20029--20038. {IEEE}, 2023.

\bibitem[\protect\citeauthoryear{Fu \bgroup \em et al.\egroup }{2023}]{abs-2307-02041}
Jie Fu, Junyu Gao, and Changsheng Xu.
\newblock Multimodal imbalance-aware gradient modulation for weakly-supervised audio-visual video parsing.
\newblock {\em CoRR}, abs/2307.02041, 2023.

\bibitem[\protect\citeauthoryear{Gao and Grauman}{2021}]{GaoG21}
Ruohan Gao and Kristen Grauman.
\newblock Visualvoice: Audio-visual speech separation with cross-modal consistency.
\newblock In {\em {CVPR}}, pages 15495--15505. Computer Vision Foundation / {IEEE}, 2021.

\bibitem[\protect\citeauthoryear{Ge \bgroup \em et al.\egroup }{2020}]{GeXWCD020}
Meng Ge, Chenglin Xu, Longbiao Wang, Eng~Siong Chng, Jianwu Dang, and Haizhou Li.
\newblock Spex+: {A} complete time domain speaker extraction network.
\newblock In {\em {INTERSPEECH}}, pages 1406--1410. {ISCA}, 2020.

\bibitem[\protect\citeauthoryear{Harte and Gillen}{2015}]{HarteG15}
Naomi Harte and Eoin Gillen.
\newblock {TCD-TIMIT:} an audio-visual corpus of continuous speech.
\newblock {\em {IEEE} Trans. Multim.}, 17(5):603--615, 2015.

\bibitem[\protect\citeauthoryear{Hsu \bgroup \em et al.\egroup }{2021}]{HsuBTLSM21}
Wei{-}Ning Hsu, Benjamin Bolte, Yao{-}Hung~Hubert Tsai, Kushal Lakhotia, Ruslan Salakhutdinov, and Abdelrahman Mohamed.
\newblock Hubert: Self-supervised speech representation learning by masked prediction of hidden units.
\newblock {\em {IEEE} {ACM} Trans. Audio Speech Lang. Process.}, 29:3451--3460, 2021.

\bibitem[\protect\citeauthoryear{Kim \bgroup \em et al.\egroup }{2021}]{KimHR21}
Minsu Kim, Joanna Hong, and Yong~Man Ro.
\newblock Lip to speech synthesis with visual context attentional {GAN}.
\newblock In {\em NeurIPS}, pages 2758--2770, 2021.

\bibitem[\protect\citeauthoryear{Kim \bgroup \em et al.\egroup }{2023}]{abs-2302-08841}
Minsu Kim, Joanna Hong, and Yong~Man Ro.
\newblock Lip-to-speech synthesis in the wild with multi-task learning.
\newblock {\em CoRR}, abs/2302.08841, 2023.

\bibitem[\protect\citeauthoryear{Lee \bgroup \em et al.\egroup }{2023}]{abs-2310-19581}
Suyeon Lee, Chaeyoung Jung, Youngjoon Jang, Jaehun Kim, and Joon~Son Chung.
\newblock Seeing through the conversation: Audio-visual speech separation based on diffusion model.
\newblock {\em CoRR}, abs/2310.19581, 2023.

\bibitem[\protect\citeauthoryear{Li \bgroup \em et al.\egroup }{2018}]{li2018effects}
Yuanqing Li, Fangyi Wang, Yongbin Chen, Andrzej Cichocki, and Terrence Sejnowski.
\newblock The effects of audiovisual inputs on solving the cocktail party problem in the human brain: An fmri study.
\newblock {\em Cerebral Cortex}, 28(10):3623--3637, 2018.

\bibitem[\protect\citeauthoryear{Li \bgroup \em et al.\egroup }{2022}]{abs-2212-10744}
Kai Li, Fenghua Xie, Hang Chen, Kexin Yuan, and Xiaolin Hu.
\newblock An audio-visual speech separation model inspired by cortico-thalamo-cortical circuits.
\newblock {\em CoRR}, abs/2212.10744, 2022.

\bibitem[\protect\citeauthoryear{Lin \bgroup \em et al.\egroup }{2023}]{abs-2306-14170}
Jiuxin Lin, Xinyu Cai, Heinrich Dinkel, Jun Chen, Zhiyong Yan, Yongqing Wang, Junbo Zhang, Zhiyong Wu, Yujun Wang, and Helen Meng.
\newblock Av-sepformer: Cross-attention sepformer for audio-visual target speaker extraction.
\newblock {\em CoRR}, abs/2306.14170, 2023.

\bibitem[\protect\citeauthoryear{Luo \bgroup \em et al.\egroup }{2023}]{abs-2306-17203}
Simian Luo, Chuanhao Yan, Chenxu Hu, and Hang Zhao.
\newblock Diff-foley: Synchronized video-to-audio synthesis with latent diffusion models.
\newblock {\em CoRR}, abs/2306.17203, 2023.

\bibitem[\protect\citeauthoryear{Lutati \bgroup \em et al.\egroup }{2023}]{abs-2301-10752}
Shahar Lutati, Eliya Nachmani, and Lior Wolf.
\newblock Separate and diffuse: Using a pretrained diffusion model for improving source separation.
\newblock {\em CoRR}, abs/2301.10752, 2023.

\bibitem[\protect\citeauthoryear{Mart{\'{\i}}nez \bgroup \em et al.\egroup }{2020}]{Martinez0PP20}
Brais Mart{\'{\i}}nez, Pingchuan Ma, Stavros Petridis, and Maja Pantic.
\newblock Lipreading using temporal convolutional networks.
\newblock In {\em {ICASSP}}, pages 6319--6323. {IEEE}, 2020.

\bibitem[\protect\citeauthoryear{Massaro and Simpson}{2014}]{massaro2014speech}
Dominic~W Massaro and Jeffry~A Simpson.
\newblock {\em Speech perception by ear and eye: A paradigm for psychological inquiry}.
\newblock Psychology Press, 2014.

\bibitem[\protect\citeauthoryear{Mu \bgroup \em et al.\egroup }{2021}]{abs-2104-09995}
Zhaoxi Mu, Xinyu Yang, and Yizhuo Dong.
\newblock Review of end-to-end speech synthesis technology based on deep learning.
\newblock {\em CoRR}, abs/2104.09995, 2021.

\bibitem[\protect\citeauthoryear{Mu \bgroup \em et al.\egroup }{2023}]{mu2023self}
Zhaoxi Mu, Xinyu Yang, Sining Sun, and Qing Yang.
\newblock Self-supervised disentangled representation learning for robust target speech extraction.
\newblock {\em CoRR}, abs/2312.10305, 2023.

\bibitem[\protect\citeauthoryear{Ochiai \bgroup \em et al.\egroup }{2019}]{OchiaiDKON19}
Tsubasa Ochiai, Marc Delcroix, Keisuke Kinoshita, Atsunori Ogawa, and Tomohiro Nakatani.
\newblock Multimodal speakerbeam: Single channel target speech extraction with audio-visual speaker clues.
\newblock In {\em {INTERSPEECH}}, pages 2718--2722. {ISCA}, 2019.

\bibitem[\protect\citeauthoryear{Pan \bgroup \em et al.\egroup }{2021}]{PanTX021}
Zexu Pan, Ruijie Tao, Chenglin Xu, and Haizhou Li.
\newblock Muse: Multi-modal target speaker extraction with visual cues.
\newblock In {\em {ICASSP}}, pages 6678--6682. {IEEE}, 2021.

\bibitem[\protect\citeauthoryear{Prajwal \bgroup \em et al.\egroup }{2020}]{PrajwalMNJ20}
K.~R. Prajwal, Rudrabha Mukhopadhyay, Vinay~P. Namboodiri, and C.~V. Jawahar.
\newblock Learning individual speaking styles for accurate lip to speech synthesis.
\newblock In {\em {CVPR}}, pages 13793--13802. Computer Vision Foundation / {IEEE}, 2020.

\bibitem[\protect\citeauthoryear{Rix \bgroup \em et al.\egroup }{2001}]{RixBHH01}
Antony~W. Rix, John~G. Beerends, Michael~P. Hollier, and Andries~P. Hekstra.
\newblock Perceptual evaluation of speech quality (pesq)-a new method for speech quality assessment of telephone networks and codecs.
\newblock In {\em {ICASSP}}, pages 749--752. {IEEE}, 2001.

\bibitem[\protect\citeauthoryear{Roux \bgroup \em et al.\egroup }{2019}]{RouxWEH19}
Jonathan~Le Roux, Scott Wisdom, Hakan Erdogan, and John~R. Hershey.
\newblock {SDR} - half-baked or well done?
\newblock In {\em {ICASSP}}, pages 626--630. {IEEE}, 2019.

\bibitem[\protect\citeauthoryear{Shi \bgroup \em et al.\egroup }{2022a}]{ShiHLM22}
Bowen Shi, Wei{-}Ning Hsu, Kushal Lakhotia, and Abdelrahman Mohamed.
\newblock Learning audio-visual speech representation by masked multimodal cluster prediction.
\newblock In {\em {ICLR}}. OpenReview.net, 2022.

\bibitem[\protect\citeauthoryear{Shi \bgroup \em et al.\egroup }{2022b}]{interspeech/ShiHM22}
Bowen Shi, Wei{-}Ning Hsu, and Abdelrahman Mohamed.
\newblock Robust self-supervised audio-visual speech recognition.
\newblock In {\em {INTERSPEECH}}, pages 2118--2122. {ISCA}, 2022.

\bibitem[\protect\citeauthoryear{Shi \bgroup \em et al.\egroup }{2022c}]{ShiHM22}
Bowen Shi, Wei{-}Ning Hsu, and Abdelrahman Mohamed.
\newblock Robust self-supervised audio-visual speech recognition.
\newblock In {\em {INTERSPEECH}}, pages 2118--2122. {ISCA}, 2022.

\bibitem[\protect\citeauthoryear{Stenzel \bgroup \em et al.\egroup }{2019}]{stenzel2019limits}
Hanne Stenzel, Jon Francombe, and Philip~JB Jackson.
\newblock Limits of perceived audio-visual spatial coherence as defined by reaction time measurements.
\newblock {\em Frontiers in neuroscience}, 13:451, 2019.

\bibitem[\protect\citeauthoryear{Subakan \bgroup \em et al.\egroup }{2021}]{SubakanRCBZ21}
Cem Subakan, Mirco Ravanelli, Samuele Cornell, Mirko Bronzi, and Jianyuan Zhong.
\newblock Attention is all you need in speech separation.
\newblock In {\em {ICASSP}}, pages 21--25. {IEEE}, 2021.

\bibitem[\protect\citeauthoryear{Vaswani \bgroup \em et al.\egroup }{2017}]{VaswaniSPUJGKP17}
Ashish Vaswani, Noam Shazeer, Niki Parmar, Jakob Uszkoreit, Llion Jones, Aidan~N. Gomez, Lukasz Kaiser, and Illia Polosukhin.
\newblock Attention is all you need.
\newblock In {\em {NIPS}}, pages 5998--6008, 2017.

\bibitem[\protect\citeauthoryear{Vincent \bgroup \em et al.\egroup }{2006}]{VincentGF06}
Emmanuel Vincent, R{\'{e}}mi Gribonval, and C{\'{e}}dric F{\'{e}}votte.
\newblock Performance measurement in blind audio source separation.
\newblock {\em {IEEE} Trans. Speech Audio Process.}, 14(4):1462--1469, 2006.

\bibitem[\protect\citeauthoryear{Wu \bgroup \em et al.\egroup }{2019}]{WuXZCYXY19}
Jian Wu, Yong Xu, Shi{-}Xiong Zhang, Lianwu Chen, Meng Yu, Lei Xie, and Dong Yu.
\newblock Time domain audio visual speech separation.
\newblock In {\em {ASRU}}, pages 667--673. {IEEE}, 2019.

\bibitem[\protect\citeauthoryear{Wu \bgroup \em et al.\egroup }{2022}]{WuJCG22}
Nan Wu, Stanislaw Jastrzebski, Kyunghyun Cho, and Krzysztof~J. Geras.
\newblock Characterizing and overcoming the greedy nature of learning in multi-modal deep neural networks.
\newblock In {\em {ICML}}, volume 162 of {\em Proceedings of Machine Learning Research}, pages 24043--24055. {PMLR}, 2022.

\bibitem[\protect\citeauthoryear{Xu \bgroup \em et al.\egroup }{2020}]{XuRCL20}
Chenglin Xu, Wei Rao, Eng~Siong Chng, and Haizhou Li.
\newblock Spex: Multi-scale time domain speaker extraction network.
\newblock {\em {IEEE} {ACM} Trans. Audio Speech Lang. Process.}, 28:1370--1384, 2020.

\bibitem[\protect\citeauthoryear{Zhou \bgroup \em et al.\egroup }{2022}]{ZhouZ0Z00O22}
Dongzhan Zhou, Xinchi Zhou, Di~Hu, Hang Zhou, Lei Bai, Ziwei Liu, and Wanli Ouyang.
\newblock Sepfusion: Finding optimal fusion structures for visual sound separation.
\newblock In {\em {AAAI}}, pages 3544--3552. {AAAI} Press, 2022.

\end{thebibliography}

\end{document}